\documentclass[12pt]{article}
\usepackage[centertags]{amsmath}
\usepackage{amsfonts}
\usepackage{amssymb}
\usepackage{graphicx}
\usepackage{amssymb}
\usepackage[verbose,colorlinks=true,naturalnames=true,linkcolor=blue,]
{hyperref}

\newcounter{rown}

\begin{document}

\title{On Twist Quantizations  of $D=4$ Lorentz and Poincar\'{e} Algebras}

\author{A. Borowiec$^{1)}$, J. Lukierski$^{1)}$ and
V.N. Tolstoy$^{1),2)}$
\\ \\
$^{1)}$Institute for Theoretical Physics,
\\University of Wroc{\l}aw, pl. Maxa Borna 9,
\\50--205 Wroc{\l}aw, Poland\\
\\$^{2)}$Institute of Nuclear Physics,
\\Moscow State University, 119 992 Moscow, Russia}

\date{$ $}
\maketitle
\begin{abstract}
We use the decomposition of $o(3,1)=sl(2;\mathbb{C})_1\oplus sl(2;\mathbb{C})_2$
in order to describe  nonstandard quantum deformation of $o(3,1)$ linked with
Jordanian deformation of $sl(2;\mathbb{C})$. Using twist
quantization technique we obtain the deformed coproducts and
antipodes which can be expressed in terms of real physical Lorentz
generators. We describe the extension of the considered deformation
of $D=4$ Lorentz algebra to the twist deformation of $D=4$
Poincar\'{e} algebra with dimensionless deformation parameter.
\end{abstract}

\section{Introduction}     
The quantum deformations of relativistic symmetries are described by
Hopf-algebraic deformation of Lorentz and  Poincar\'{e} algebras.
Such quantum deformations are classified by Lorentz and Poincar\'{e}
Poisson structures. These Poisson structures given by classical
$r$-matrices were classified already some time ago by S. Zakrzewski
(see \cite{blt1} for the Lorentz classical $r$-matrices and
\cite{blt2} for  the Poincar\'{e} classical matrices). In
\cite{blt1} there are provided four classical $o(3,1)$ $r$-matrices
and in \cite{blt2} one finds 21 cases describing different
deformation of Poincar\'{e} symmetries, with various numbers of free
parameters.

In this paper we would like to describe the explicit Hopf algebra
form of the nonstandard deformations of $D=4$ Lorentz algebra and
extend it to $D=4$ Poincar\'{e} algebra. We shall describe firstly
the twist deformation of complexified $D=4$ Lorentz algebra
$o(4;\mathbb{C})$, and further consider the reality structure
($*-$Hopf algebra) describing quantum deformation of the standard
Lorentz algebra $o(3,1)$.
Let us observe that the complex Lie algebra $o(4;\mathbb{C})$
decomposes into a direct sum of two copies
of $o(3;\mathbb{C})\approx sl(2;\mathbb{C})$ algebras.
We shall employ the $sl(2;\mathbb{C})_1\oplus sl(2;\mathbb{C})_2$
basis $(H_1,\,E_1,\,F_1)\oplus ( H_2,\, E_2,\, F_2)$ where
\begin{equation}\label{blte1}
[H_k,\,E_k]=E_k\, \qquad [H_k,\,F_k]=-F_k\, \qquad [E_k,\,F_k]=2H_k \, ,\qquad k=1,2
\end{equation}
and obviously
\begin{equation}\label{blte2}
[X_1,\, X_2]=0,\, \qquad X\in (H,\,E,\,F)
\end{equation}
The real $o(1,3)$ Lorentz algebra is obtained from the reality condition
$X_1^\star =-X_2$ or equivalently $X_2^\star =-X_1$.
The real generators ($x^\star =-x$, for $x\in (x, x')$), where
 $X_{1,2}=x\pm i x'$ \footnote{We see that besides of the
$\star$-operation one can also introduce a  standard complex
conjugation operation by $\overline{X_1} =X_2 $ and $\overline{X_2}
=X_1$.} we describe explicitly  as follows
\begin{eqnarray}\label{blte3}
   h=\frac{1}{2}(H_1+H_2), \, \qquad  e=\frac{1}{2}(E_1+E_2), \,
   \qquad  f=\frac{1}{ 2}(F_1+F_2) \, \cr\cr
   h'=\frac{-i}{2}(H_1-H_2), \, \qquad  e'=\frac{-i}{2}(E_1-E_2), \,
   \qquad  f'=\frac{-i}{ 2}(F_1-F_2) \,
\end{eqnarray}
Notice that these new generators satisfy the commutation relation of
the Lorentz algebra $o(1,3)$ written in Cartan-Chevaley basis:
\begin{eqnarray}\label{blte4}
&   [h, e] = e \, , \qquad [h,f] = - f \, ,\qquad [e,f] = 2h
   \cr \cr
 &  [h,e']=[h',e]=e'\,,\qquad [h',f]=[h,f']=-f'\,,\qquad  [e,f']=[e',f] =2h' \, ,
\cr    \cr
  &  [h',e'] = -e \, , \qquad [h', f']= f \, , \qquad [e',f'] = - 2h
\end{eqnarray}
with vanishing all remaining relations.

We recall that the quantization of classical Lie-algebra $g$ is obtained by
introducing the twist function ${\mathcal F} \in U(g) \otimes
U(g)$ which modifies the coproduct $\Delta$ and antipode $S$
as follows \cite{blt10}:
\begin{equation}\label{blte7}
    \Delta \longrightarrow \Delta_{{\mathcal F}} = {\mathcal F} \, \Delta^{(0)} \,
    {\mathcal F}^{-1} \, ,
    \qquad
    S \longrightarrow S_{\mathcal F} = u \, S \, u^{-1} \, ,
\end{equation}
where
\begin{eqnarray}\label{blte8}
&\Delta^{(0)} (x) =  x \otimes 1 + 1 \otimes x,\ \ \ \ \forall x\in g
 \cr\cr
&  {\mathcal F}  = \sum\limits_{i} f^{(1)}_{i} \otimes  f^{(2)}_{i} \, , \qquad u =
\sum\limits_{i}  f^{(1)}_{i}\, S\, ( f^{(2)}_{i})\, .
\end{eqnarray}
It appears that in classical enveloping Lie algebra $U(g)$, considered as a
Hopf algebra $H^{(0)} = (U (g), m, \Delta^{(0)}, S, \epsilon)$ only coalgebra sector
(coproduct and coinverse) is modified.

The standard Drinfeld-Jimbo quantum deformation of $D=4$  Lorentz algebra
(see \cite{blt4,blt5})
satisfies modified YB equation and can not be extended to Poincare algebra. In
this paper we shall consider the nonstandard quantum deformations satisfying CYBE
which can be extended to the whole Poincar\'{e}
algebra and provide new deformation of relativistic symmetries.
We shall consider here more in detail the two-parameter nonstandard deformation generated by  
the classical $sl(2;\mathbb{C})$ $r$-matrix $r(\alpha,\beta)$ 
\begin{eqnarray}\label{blte9}
 r(\alpha,\beta) &=& \alpha (h \wedge e-h' \wedge e') +
    \beta \, e \wedge e' \, \nonumber\\
    \ & =& \alpha (H_1 \wedge E_1 + H_2 \wedge E_2 )+i\beta\, E_1\wedge E_2
      \, ,
\end{eqnarray}
Further we shall assume that the real $r$-matrix (\ref{blte9}) is anti-Hermitean, i.e.
$H_1^*=-H_2,\ E_1^*=-E_2$ and the  parameters $\alpha$ and $\beta$ should be purely imaginary.

We see that for $\beta=0$ the $r$-matrix (\ref{blte9}) describes the real part of complex Jordanian $r$-matrix for $sl(2,\mathbb{C})\approx o(3;\mathbb{C})$, which is quantized by the
following Ogievetsky twist ($k=1,2$) (see \cite{blt8})
\begin{equation}\label{blte10}
    {\mathcal F}_{J,k} = \exp{(H_k\otimes \Sigma_k)} \,
\end{equation}
 where $\Sigma_k=ln{(1+\alpha E_k)}$. Because the generators $(H_1,\, E_1)$ and $(H_2,\, E_2)$ do commute
 the twist function corresponding
 to (\ref{blte9}) is given by the following formula:
 \begin{equation}\label{blte11}
    {\mathcal F}(\alpha,\beta)= {\mathcal F}_R {\mathcal F}_{J,1}{\mathcal F}_{J,2}  \,
\end{equation}
where
\begin{equation}\label{blte12}
{\mathcal F}_R=\exp{(i\,\beta\Sigma_1\wedge\Sigma_2)}
\end{equation}
We would like to mention here that the form of the twist function
given above  by formula (\ref{blte10}) was conjecture with not antisymmetrized
exponential form (\ref{blte12})
by Kulish and Mudrov in \cite{KM}.

In order to deform the real Lorentz algebra we demand that our twist
quantization is compatible with $\star$-algebra structure. More
explicitly, this means that the twist ${\mathcal F}(\alpha,\beta)$
must be $\star$-unitary, i.e. ${\mathcal
F}(\alpha,\beta)^\star={\mathcal F}(\alpha,\beta)^{-1}$. As we have
mentioned above the last equation can be statisfied if and only if
both parameters $\alpha,\,\beta$ are purely imaginary
($\Sigma_1^\star=\Sigma_2$)\footnote{However
$\overline{\Sigma_1}\neq \Sigma_2$.}, i.e. one has
\begin{equation}\label{blte13}
 ({\mathcal F}_{J,k})^\star={\mathcal F}_{J,k+1}^{-1}  ,\qquad
 ({\mathcal F}_R)^\star={\mathcal F}_R^{-1}\,.
\end{equation}

In Sect. 2 we shall describe the deformed coproducts in the
classical $sl(2;\mathbb{C})_1\oplus sl(2;\mathbb{C})_2$ basis
$(H_1,\, E_1,\,F_1)\oplus (H_2,\,E_2,\,F_2)$ which can be expressed
easily  in terms of physical (Hermitean) $o(3,1)$ Lorentz algebra
basis $(M_i,\, N_i)$, satisfying the algebra
\begin{equation}\label{blte14}
    [M_i,M_j]=i\epsilon_{ijk}M_k,\, \qquad [M_i,N_j]=i\epsilon_{ijk}N_k,\, \qquad [N_i,N_j]=-i\epsilon_{ijk}M_k,
\end{equation}
where using (\ref{blte3}) one can find that
\begin{eqnarray}\label{blte29}
 h=i\,N_3, \qquad h'=-i\,M_3\nonumber\\
 e=i\,(N_1+\,M_2), \qquad e'=i\,(N_2-\,M_1)\\
 f=i\,(N_1-\,M_2), \qquad f'=-i\,(N_2+\,M_1)\nonumber
 \end{eqnarray}
 In Sect. 3 we shall add to the Lorentz algebra (\ref{blte14}) four momentum generators and describe
 the quantum deformation of Poincar\'{e} algebra,
 generated by classical $r$-matrix (\ref{blte9}) (or equivalently (\ref{blte11})).
Finally in Sect. 4 we shall present an outlook and mention possible applications.

 \section{Two-parameter nonstandard deformation of $o(3,1)$ in classical basis}

If we use (\ref{blte1}-\ref{blte2}), (\ref{blte7}) and (\ref{blte11}) we obtain the following formulae for the coproducts of $sl(2;\mathbb{C})\approx o(3;\mathbb{C})$ generators  $(H_k,\, E_k,\,F_k)$, $k=1,2$
\begin{eqnarray}\label{blte21}
\Delta_{\alpha,\beta}(E_k)
&=&{\mathcal
F}(\alpha,\beta)\Delta^{(0)}(E_k)
{\mathcal F}^{-1}(\alpha,\beta)=\Delta_\alpha(E_k)=E_k\otimes e^{\Sigma_k}+1\otimes E_k
\nonumber
\\
&& \\
\Delta_{\alpha,\beta}(H_k)
&= & H _k\otimes e^{-\Sigma_k}+1\otimes H_k-(-1)^ki\beta\Lambda_k e^{\Sigma_k}\otimes\Sigma_{k+1} e^{-\Sigma_k} 
\nonumber \\[8pt]&&
+\,(-1)^ki\beta\Sigma_{k+1}\otimes\Lambda_k e^{\Sigma_k}
\nonumber
\\
&& \\
\Delta_{\alpha,\beta}(F_k)&=&F_k\otimes e^{-\Sigma_k}+1\otimes F_k
+2\alpha H_k\otimes H_ke^{-\Sigma_k}
 +\alpha H_k(H_k-1)\otimes\Lambda_k+
\nonumber \\[8pt]&&
2(-)^ki\alpha\beta H_k\Sigma_{k+1}\otimes\Lambda_k
 -2(-)^ki\alpha\beta
H_k\otimes\Sigma_{k+1}e^{-\Sigma_k}
\nonumber \\[8pt]&&
-\,2(-)^ki\alpha\beta H_k\Lambda_k
e^{\Sigma_k}\otimes\Sigma_{k+1}e^{-2\Sigma_k}
 -\,2(-)^ki\alpha\beta\Lambda_k e^{\Sigma_k}\otimes
H_k\Sigma_{k+1}e^{-\Sigma_k}\,
\nonumber \\[8pt]&&
 +\,2(-)^ki\alpha\beta\Sigma_{k+1}\otimes H_ke^{-\Sigma_k}
 +\,(-)^ki\alpha\beta\Lambda_k\otimes\Sigma_{k+1}e^{-\Sigma_k}\,-
 \nonumber \\[8pt]&&
 (-)^k\,i\alpha\beta\Sigma_{k+1}\otimes\Lambda_k+(-)^ki\alpha\beta(e^{-2\Sigma_k}-1)
\otimes\Sigma_{k+1}\Lambda_k
 \nonumber
\\[8pt]
&&
-\,\alpha\beta^2\Lambda_k\otimes\Sigma^2_{k+1}e^{-\Sigma_k}
-\alpha\beta^2\Sigma^2_{k+1}\otimes\Lambda_k
-\alpha\beta^2\Lambda^2_ke^{2\Sigma_{k}}\otimes\Sigma^2_{k+1}\Lambda_k\ 
 \nonumber\\[8pt]
&&
 \end{eqnarray}
Here $\Sigma_{k+1}$ is understood with index mod 2, i.e. $\Sigma_{k+1}$ denotes
$\Sigma_1$ for $k=2$ and $\Lambda_k=e^{-2\Sigma_k}-e^{-\Sigma_k}$.

Using the relations (\ref{blte7}-\ref{blte8}) one obtains the following formulae for the antipodes
 \begin{eqnarray}\label{blte22}
  S_{\alpha,\beta}(E_k)&=& S_{\alpha}(E_k)=-E_ke^{-\Sigma_k}  ,\qquad  S_{\alpha,\beta}(H_k)= S_{\alpha}(H_k)=-H_ke^{\Sigma_k}\\[8pt]
  S_{\alpha,\beta}(F_k)&=&-F_ke^{\Sigma_k}
  +\alpha H^2_ke^{\Sigma_k}(e^{\Sigma_k}+1)
  -\alpha^2 H_kE_ke^{\Sigma_k}
  +2\alpha^3\beta^2\Sigma_{k+1}^2E_k^2\ \ 
\end{eqnarray}

In order to apply our results to the physical basis of the Lorentz algebra one has to calculate the coproducts
for the generators $(h,h',e,e',f,f')$  by using (\ref{blte3}), (\ref{blte21})-(16)
 and further the formulae (\ref{blte29}).

\section{Extension of  
the deformation to Poincar\'{e} algebra}

The $D=4$ Lorentz algebra (\ref{blte14}) can be extended to $D=4$
Poincar\'{e} algebra by adding the mutually commuting
four-momentum operators ($P_{0},\,P_{1},\,P_{2},\,P_{3})$
satisfying the relations ($j,k,l=1,2,3$)
\begin{equation}
\begin{array}{rcl}
[M_{j},\,P_{l}]\!\!& =\!\!&i\,\epsilon_{jlk}\,P_{k}~,\qquad\;
[M_{j},\,P_{0}]\;=\;0~,
\\[7pt]
[N_{j},\,P_{k}]\!\!&=\!\!&-i\,\delta_{jk}\,P_{0}~,\qquad
[N_{j},\,P_{0}]\;=\;-i\,P_{j}~.
\end{array}\label{dp1}
\end{equation}
Because  the classical $r-$matrix (\ref{blte9}) for $D=4$ Lorentz algebra satisfies CYBE
it provides also the deformation of the $D=4$ Poincar\'{e} algebra. The twisted coproducts
of the four-momenta in the "spinorial" basis $P_\pm=P_{0}\pm P_{3}$, $P'_\pm=P_{1}\pm iP_{2}$
are given by the formulae:
\begin{eqnarray}\label{blteP}
\Delta_{\alpha,\beta}(P_+)
&=&{\mathcal
F}(\alpha,\beta)\Delta^{(0)}(P_+)
{\mathcal F}^{-1}(\alpha,\beta)=\Delta_\alpha(P_+)=P_+\otimes
e^{\frac{1}{2}(\Sigma_1+\Sigma_2)}+1\otimes P_+
\nonumber
\\
&& \\
\Delta_{\alpha,\beta}(P'_-)
&= & P'_-\otimes e^{\frac{1}{2}(\Sigma_2-\Sigma_1)}+1\otimes P'_-
+\alpha H_1\otimes P_+ e^{-\Sigma_1}\,+\nonumber \\[8pt]&&
i\alpha\beta P_+e^{-\Sigma_1}\otimes \Sigma_{2}e^{\frac{1}{2}(\Sigma_2-\Sigma_1)}
+i\alpha\beta\Lambda_1 e^{\Sigma_1}\otimes\Sigma_{2}P_+ e^{-\Sigma_1} -i\alpha\beta\Sigma_{2}\otimes P_+ e^{-\Sigma_1}
\nonumber
\\
&& \\
\Delta_{\alpha,\beta}(P'_+)
&= & P'_+\otimes e^{\frac{1}{2}(\Sigma_1-\Sigma_2)}+1\otimes P'_+
+\alpha H_2\otimes P_+ e^{-\Sigma_2}\,-\nonumber \\[8pt]&&
i\alpha\beta P_+e^{-\Sigma_2}\otimes \Sigma_{1}e^{\frac{1}{2}(\Sigma_1-\Sigma_2)}
-i\alpha\beta\Lambda_2 e^{\Sigma_2}\otimes\Sigma_{1}P_+ e^{-\Sigma_2} +i\alpha\beta\Sigma_{1}\otimes P_+ e^{-\Sigma_2}
\nonumber
\\
&& \\
\Delta_{\alpha,\beta}(P_-)&=&P_-\otimes e^{-\frac{1}{2}(\Sigma_1+\Sigma_2)}+
1\otimes P_-\,+\alpha P'_+e^{-\Sigma_1}\otimes \Sigma_2 e^{-\frac{1}{2}(\Sigma_1+\Sigma_2)}+
\nonumber \\[8pt]&&
+\alpha H_1\otimes P'_+e^{-\Sigma_1}+\alpha H_2\otimes P'_-e^{-\Sigma_2}
+\alpha^2 H_1H_2\otimes P_+e^{-(\Sigma_1+\Sigma_2)}+
\nonumber \\[8pt]&&
-i\alpha\beta P'_-e^{-\Sigma_2}\otimes \Sigma_1 e^{-\frac{1}{2}(\Sigma_1+\Sigma_2)}
+i\alpha\beta \Sigma_1\otimes P'_- e^{-\Sigma_2}
+i\alpha\beta \Lambda_1 e^{\Sigma_1}\otimes \Sigma_2P'_- e^{-\Sigma_1}
\nonumber \\[8pt]&&
-i\alpha\beta \Sigma_2\otimes P'_+ e^{-\Sigma_1}
-i\alpha\beta \Lambda_2e^{\Sigma_2}\otimes \Sigma_1P'_- e^{-\Sigma_2}
+i\alpha^2\beta \Sigma_{1}H_1\otimes P_+e^{-(\Sigma_1+\Sigma_2)}
\nonumber \\[8pt]&&
 +i\alpha^2\beta H_2\Lambda_1e^{\Sigma_1}\otimes\Sigma_{2}P_+e^{-(\Sigma_1+\Sigma_2)}
 -i\alpha^2\beta P_+e^{-(\Sigma_1+\Sigma_2)}\otimes \Sigma_1\Sigma_2 e^{-\frac{1}{2}(\Sigma_1+\Sigma_2)}
\nonumber \\[8pt]&&
-i\alpha^2\beta \Sigma_{2}H_2\otimes P_+e^{-(\Sigma_1+\Sigma_2)}
-i\alpha^2\beta H_1\Lambda_2e^{\Sigma_2}\otimes\Sigma_{1}P_+e^{-(\Sigma_1+\Sigma_2)}
\nonumber
\\[8pt]
&&
+\alpha^2\beta^2\Sigma_1\Sigma_{2}\otimes P_+e^{-(\Sigma_1+\Sigma_2)}
+\alpha^2\beta^2\Lambda_1\Lambda_{2}e^{(\Sigma_1+\Sigma_2)}\otimes \Sigma_1\Sigma_{2}P_+e^{-(\Sigma_1+\Sigma_2)}
 \nonumber
\\[8pt]
&&
 \end{eqnarray}
 Let us observe that in the purely Jordanian case ($\beta=0$) all formulae
 (14)-(16) and (20)-(23) simplify significantly.

 \section{Outlook}
 The description in detail of all possible Hopf algebra structures of $D=4$ quantum Lorentz algebras should
 have a physical importance. In particular the nonstandard deformations of the Lorentz algebra described by classical $r-$matrices satisfying CYBE are described by twist quantization and can be extended to the Poincar\'{e} algebra. Recently it has been shown that that the twist
 quantizations of Poincar\'{e} algebra is very useful in the construction of new
 noncommutative deformed Minkowski spaces which are covariant under relativistic symmetries
 see (e.g. \cite{jwess}-\cite{LW}).

One can further observe that $o(3,1)$ describes as well the $D=3$ de-Sitter algebra, which provides 
via quantum AdS contraction the $D=3$ deformed Poincar\'{e} algebra.  These novel aspects of the presented 
deformation as well as the structure of
 dual deformed Lorentz and Poincar\'{e} groups will be considered in our
 next publication.

\bigskip
{\small Ackowledgement:
A.B. and J.L. are supported by KBN grant 1PO3BO1828.
V.N.T. is supported by the Russian Foundations for Fundamental Research grants
RFBR-05-01-01086 and INTAS OPEN 03-51-3350.}

\end{document}